\documentstyle[11pt,psfig,fullpage,citesort]{article}

\def\Tr{{\rm Tr}}
\newcommand{\ket}[1]{|#1\rangle}

\def\m@th{\mathsurround=0pt }
\def\leftrightarrowfill{$\m@th \mathord\leftarrow \mkern-6mu
 \cleaders\hbox{$\mkern-2mu \mathord- \mkern-2mu$}\hfill
 \mkern-6mu \mathord\rightarrow$}
\def\overleftrightarrow#1{\vbox{\ialign{##\crcr
     \leftrightarrowfill\crcr\noalign{\kern-1pt\nointerlineskip}
     $\hfil\displaystyle{#1}\hfil$\crcr}}}

\begin{document}
\renewcommand{\thefootnote}{\fnsymbol{footnote}}
\begin{titlepage}
\begin{flushright}
UFIFT-HEP-00-32\\
hep-th/0012006
\end{flushright}

\vskip 1.5cm

\begin{center}
\begin{Large}
{\bf A Six Vertex Model on a Fishnet}
\end{Large}

\vskip 2.cm

{\large Charles B. Thorn\footnote{E-mail  address: thorn@phys.ufl.edu}}

\vskip 0.5cm

{\it Institute for Fundamental Theory\\
Department of Physics, University of Florida,
Gainesville, FL 32611}

\vskip .5cm
\end{center}

\begin{abstract}
\noindent 
The flow of $U(1)$ charge through dense fishnet diagrams,
in a non-hermitian matrix scalar field theory
$g_1\Tr(\Sigma^\dagger\Sigma)^2+2g_1v\Tr\Sigma^{\dagger2}\Sigma^2$, 
is described by a 6-vertex model on a
``diamond'' lattice \cite{gilesmt}.
We give a direct calculation of the
continuum properties of the 6-vertex model on this
novel lattice, explicitly confirming the conclusions of \cite{gilesmt},
that, for $1/2\leq v<\infty$, they are identical to those of a world-sheet
scalar field compactified on a circle $S_1$.
The radius of the circle is related to the ratio $v$ of 
quartic couplings by $R^{-2}=2T_0\cos^{-1}(1-1/2v^2)$. 
This direct computational approach may be of value in 
generalizing the conclusions to the non-Abelian $O(n)$ case.
\end{abstract}
\end{titlepage}
\section{Introduction}
We are accustomed in string and related theories to the 
apparent need for more than four space-time dimensions. On
the other hand, the conjectured equivalence (duality), between
some ordinary four dimensional supersymmetric quantum field theories and
string theories on certain ten dimensional space-time manifolds 
\cite{maldacena}, suggests that
these extra dimensions may be but a mathematical device that
conveniently reflects some other aspect of the underlying theory.

In the late 1970's, we showed how a single extra compact dimension
effectively arises in the context of the fishnet diagram model
of a string world-sheet in a scalar quantum field theory with
an $O(2)\equiv U(1)$ global symmetry \cite{gilesmt}. It is
plausible that in field theories with a richer symmetry, for instance
$SO(n)$, multiple compact dimensions (perhaps $S_{n-1}$) will arise from
the same mechanism. 

As an important preliminary to extending the work of
\cite{gilesmt} to higher symmetries, we give in this article a 
new derivation of the main
results of \cite{gilesmt} that doesn't
rely on indirect arguments, such as invoking $\sigma\leftrightarrow\tau$ 
symmetry (modular invariance)
and the universality of the continuum limit. Instead, we
obtain all the results 
by direct and explicit computation. The methods
used here are fairly standard, borrowed from the mathematical
analysis of one-dimensional Heisenberg spin chains by Yang and
Yang \cite{yangyang} and from that of the 6-vertex models
by Lieb \cite{lieb} and others \cite{liebmw}. 
However our particular model involves several
novel details, e.g. a non-standard lattice, 
not present in those classic treatments,
and we think our direct approach illuminates these features
in a useful way. 

\begin{figure}[ht]
\centerline{\psfig{file=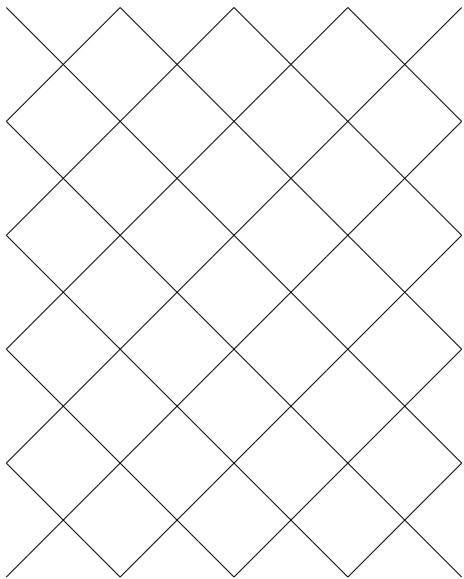,height=3.0in}}
\caption[]{Fishnet diagram for a scalar quantum field theory
with only quartic interactions.}
\label{fishgen}
\end{figure}

The fishnet model \cite{nielsenfishnet} as formulated in
\cite{thornfishnet} is obtained in the context of a 
discretized light-cone parameterization of the propagators
of an $N_c\times N_c$ matrix quantum field theory:
\begin{eqnarray}
D({\bf p}, p^+, x^+)&=&\theta(\tau){1\over 2p^+}
e^{-\tau{\bf p}^2/2p^+}\to{1\over 2l}
e^{-k{\bf p}^2/2lT_0}
\end{eqnarray} 
where $p^+=lm$ and $\tau\equiv ix^+=ka\equiv km/T_0$ 
with $k,l$ running over all positive integers. Planar
diagrams are singled out by 't Hooft's $N_c\to\infty$
limit. All propagators of the fishnet diagrams
(see Fig.~\ref{fishgen}) of \cite{thornfishnet}
are restricted to $k=l=1$.
For a fixed total $P^+=Mm$ and
fixed evolution time $T=Na$, this restriction achieves the maximal
density of vertices, and is plausibly appropriate for 
strong coupling. In \cite{gilesmt} and also here, the focus is on
how to deal with internal degrees of freedom carried
by the propagators of these fishnet diagrams. We simply
accept the simplest fishnet {\it ansatz} as a working model
for string formation. As discussed in \cite{beringrt}
there is a lot of room for improvements in the fishnet model
itself, but we don't consider those issues here.

Thus the sum of fishnet diagrams provides a model of the discretized 
world-sheet \cite{gilest} of light-cone string theory \cite{goddardgrt}. 
The main piece of physics we draw from the model is that
$M$ and $N$, which determine the size and shape of the lattice
world-sheet,
are determined by the $P^+$, carried by the system of field
quanta propagated by the diagram, and by
the time span over which that system evolves. In the continuum 
limit $N/M\to T/P^+T_0$, so that energy ($P^-$) eigenvalues $E_r$ 
can be read off
from the exponential dependence on $N$ of the sum over diagrams
$\sim e^{-Na E_r}$.

Notice that the fishnet diagram determines a novel two
dimensional square lattice in which all links are rotated
by 45 degrees. We refer to such a lattice as a diamond 
lattice. One of the lacunae in \cite{gilesmt} was the reliance on
known results  for a conventional square
lattice plus the reasonable assertion that the continuum limit should be isotropic.
Our calculation here deals directly with the diamond
lattice configuration, and the details of the computation
are therefore new. In particular we develop an appropriate
adaptation of the Bethe {\it ansatz} \cite{bethe} to the diamond lattice.

\section{The Model}
\begin{figure}[ht]
\centerline{\psfig{file=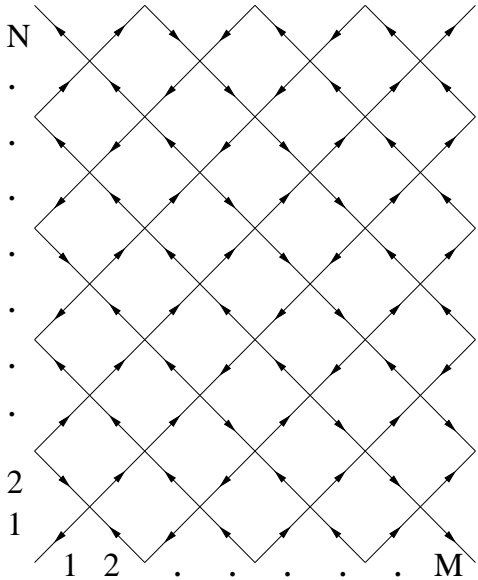,height=3.0in}}
\caption[]{Fishnet propagating $M$ units of $P^+$ $N$ steps
in time. Periodic boundary conditions have been imposed.}
\label{6vfishnet}
\end{figure}

The field theory analyzed in \cite{gilesmt} is the special
case $n=2$ of a matrix scalar field ${\Sigma_i}$
transforming as 
a vector under $O(n)$ with quartic interaction terms that respect this symmetry:
\begin{eqnarray}
{\cal L}&=& -{1\over2}\Tr(\partial\Sigma_i)^2 -{\mu^2\over2}\Tr\Sigma_i^2
-{g_1\over2}\Tr\Sigma_i^2\Sigma_j^2
-{g_2\over4}\Tr\Sigma_i\Sigma_j\Sigma_i\Sigma_j.
\end{eqnarray}  
For $n=2$ it is convenient  to
define a single non-hermitian 
matrix field $\Sigma\equiv(\Sigma_1+i\Sigma_2)/\sqrt2$
which carries a unit $U(1)\equiv SO(2)$ charge.
The interaction terms then become
\begin{eqnarray}
V(\Sigma)=(g_1+g_2)\Tr\Sigma^{\dagger2}\Sigma^2
+g_1\Tr\Sigma^\dagger\Sigma\Sigma^\dagger\Sigma.
\end{eqnarray}
The charge carried by a given line in a diagram is indicated by
attaching an arrow pointing in the direction of charge flow.
There are precisely
six (planar) charge conserving vertices (see Fig.~\ref{6vertices}): 
\begin{figure}[ht]
\centerline{\psfig{file=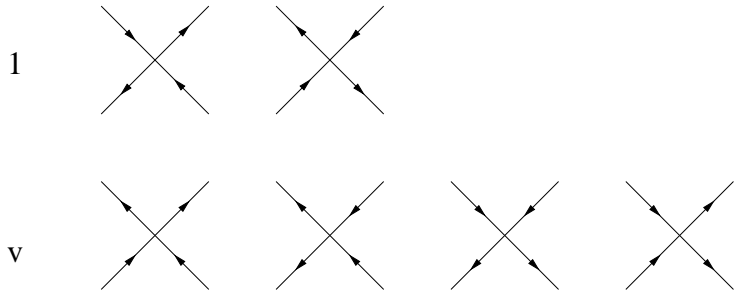,height=1.5in}}
\caption[]{$U(1)$ Vertices: $v=(g_1+g_2)/2g_1$}
\label{6vertices}
\end{figure}
Two with weight $2g_1$ 
in which each adjacent
pair carries charge 0 into the vertex, and four with weight
$g_1+g_2\equiv 2vg_1$ in which two adjacent
lines carry charge 2 into the vertex. Scaling $g_1, g_2\to \lambda g_1,
\lambda g_2$ just multiplies the diagram by an overall factor
$\lambda^{MN}$, so we lose nothing by scaling $2g_1$ to 1, so the
weights of the two classes of vertices are $1, v$ respectively.
A typical fishnet diagram with these vertices is shown in
Fig.~\ref{6vfishnet}. The sum of all such diagrams is
thus seen to be equivalent to calculating the 
partition function for a 6-vertex model on a diamond
lattice.
\section{The transfer Matrix and its Eigenvalues}
The fishnet diagram can be thought of as a discrete (imaginary) time
evolution of a state which is a tensor product of $M$ two state systems,
(``spins''), labeled by up and down arrows. Because of the
diamond lattice configuration, the basic discrete evolution is two
time steps, and we define each element of the
$2^M\times2^M$ transfer matrix ${\cal T}$ as the product of
vertex factors associated with the subgraph that connects
a given row of arrows with the row two time steps above it.
It is easy to see that the state with all arrows up or all arrows
down is an eigenstate of the transfer matrix with
eigenvalue $v^M$.
\subsection{One overturned arrow: $Q=M-2$}
Let us next consider the states with
one overturned arrow, i.e. $M-1$ up arrows and
1 down arrow. With periodic boundary conditions, even and odd locations 
for this down arrow are not equivalent, because of the diamond lattice.
Denote the state with down arrow at location $j$ by $\ket{j}$.
Then, by following the change in $j$ after two time steps we
find the action of the transfer matrix
\begin{eqnarray}
{\cal T}\ket{j}=\cases{\ket{j+2}v^M+\ket{j+1}v^{M-1}+\ket{j-1}v^{M-1}
+\ket{j}v^{M-2}& for $j$ odd\cr
\ket{j-2}v^M+\ket{j+1}v^{M-1}+\ket{j-1}v^{M-1}
+\ket{j}v^{M-2}& for $j$ even.\cr}
\end{eqnarray}
Note, by the way, that the action of ${\cal T}$ is local in that the
down spin migrates at most two sites after two time steps.\footnote{In contrast,
for the 6-vertex model on a conventional square lattice, the down spin
can be on any site after a single time step. This is one sense in which the
diamond lattice is superior from a physical point of view.}
Thus, in representing a spin wave as a Fourier transform with respect
to location, we must allow for a phase shift in the even terms with respect
to the odd terms:
\begin{eqnarray}
\ket{k}=\sum_{j~{\rm odd}}\ket{j}e^{ikj}
+\xi(k)\sum_{j~{\rm even}}\ket{j}e^{ikj}
\end{eqnarray}
Applying the transfer matrix to the state $\ket{k}$ and shifting
the sums over $j$ appropriately, we find that $\ket{j}e^{ikj}$
is multiplied by the factor
\begin{eqnarray}
v^{M-2}+e^{-2ik}v^M+\xi(k)e^{ik}v^{M-1}
+\xi(k)e^{-ik}v^{M-1}\quad &&{\rm for}~j~{\rm odd}\nonumber\\
\xi(k)v^{M-2}+\xi(k)e^{2ik}v^M+e^{ik}v^{M-1}
+e^{-ik}v^{M-1}\quad &&{\rm for}~j~{\rm even},
\label{factors}
\end{eqnarray}
for $j\neq 1,M$. For $\ket{k}$ to be an eigenstate of ${\cal T}$, these two
factors must agree, which determines a quadratic equation for
$\xi(k)$:
\begin{eqnarray}
\xi^2-(2iv\sin k)\xi-1=0.
\end{eqnarray}
It is easy to see that interchanging the two solutions for $\xi$
interchanges the eigenstate $\ket{k}$ with $-\ket{k+\pi}$. Thus
we lose no generality in selecting the solution
\begin{eqnarray}
\xi(k)&\equiv&iv\sin k+\sqrt{1-v^2\sin^2 k}
\end{eqnarray}
provided we allow $k$ the full $2\pi$ range, $-\pi<k\leq\pi$. 
(Keeping both solutions would require restricting the range of $k$
to $\pi$ to avoid double counting.)
When the transfer matrix acts on arrows near $j=1,M$, it wraps
around with the location 1 equivalent to $M+1$, $M$ equivalent
to 0, etc. These terms must match for $\ket{k}$ to be an
eigenstate, which requires $e^{iMk}=1$ or
\begin{eqnarray}
k&=&{2\pi n\over M}\qquad {\rm for}~n=0,1,2,\ldots, M-1.
\end{eqnarray}
Putting the selected solution for $\xi(k)$ back into one of the factors
(\ref{factors}), we find the eigenvalue of the transfer matrix to be
$v^Mt(k)$ where
\begin{eqnarray}
t(k)&=&\left(\cos k+{1\over v}\sqrt{1-v^2\sin^2 k}\right)^2.
\end{eqnarray}
It is a welcome outcome that the eigenvalue is automatically
positive as long as $v^2\sin^2 k<1$.
\subsection{Two or more overturned arrows: $Q=M-2q$} 
Moving on to two overturned arrows, we employ
the Bethe {\it ansatz} for two overturned arrows appropriate to
our diamond lattice:
\begin{eqnarray}
\ket{k_1,k_2}&=&\left[\sum_{l~{\rm odd}}\sum_{m~{\rm odd}}
+\xi_1\sum_{l~{\rm even}}\sum_{m~{\rm odd}}
+\xi_2\sum_{l~{\rm odd}}\sum_{m~{\rm even}}
+\xi_1\xi_2\sum_{l~{\rm even}}\sum_{m~{\rm even}}\right]
\ket{l,m}e^{ilk_1+imk_2}\nonumber\\
&&\nonumber\\
&&\hskip2in\mbox{}+A(1,2)\Bigl(k_1,\xi_1
\leftrightarrow k_2,\xi_2\Bigr)
\end{eqnarray}
For $q$ overturned arrows, the Bethe {\it ansatz} is the obvious
generalization, wherein the sum over $1\leftrightarrow2$ is
replaced by a sum over all permutations of the spins, and
$A(1,2)$ is generalized to an $A_P$ for each permutation.
$A_P$ factors into a product of $A(k,l)$ for each pair interchange
needed to accomplish the permutation. When the Bethe {\it ansatz}
succeeds, as it does for this model, the eigenvalue of the transfer matrix
is just
\begin{eqnarray}
T(k_1,\ldots,k_q)=v^M\prod_{j=1}^q 
t(k_j).
\end{eqnarray}
By explicitly analyzing the case $q=2$, we find after much tedious algebra that
\begin{eqnarray}
A(1,2)=-{(1-1/v^2)z_2-z_1-z_1z_2/v-1/v\over
    (1-1/v^2)z_1-z_2-z_1z_2/v-1/v},
\end{eqnarray}
where we have defined $z_j\equiv \xi(k_j)e^{ik_j}$. Finally, the
conditions that the periodic boundary conditions are consistent
with the {\it ansatz} being an eigenstate are
\begin{eqnarray}
e^{iMk_2}=A(1,2), \qquad e^{iMk_1}=A(2,1)=1/A(1,2).
\end{eqnarray}
Since the
permutation factor $A_P$ for $q$ overturned arrows is built
up from pair factors, the $q=2$ case is sufficient to determine all
of the information we need to handle the general case, which
leads to the boundary conditions
\begin{eqnarray}
e^{iMk_l}=\prod_{j\neq l}A(j,l).
\end{eqnarray}
\section{Analysis of the continuum limit: $M,N\to\infty$}
For analyzing these equations it is convenient \cite{yangyang} to map the $k_j$
onto new variables $\alpha_j$ for which $A(j,l)$ depends only on the
difference $\alpha_j-\alpha_l$. This is accomplished by the map
\begin{eqnarray}
z&=&\xi e^{ik}={e^{i\nu}-e^\alpha\over e^{i\nu+\alpha}-1}\nonumber\\
e^{i\nu}&=&{1\over 2v}+i\sqrt{1-{1\over4v^2}}.
\end{eqnarray}
Note that our parameter $\nu$ is related to a similar parameter $\mu$
in \cite{yangyang} by $\mu=2\nu$, which we shall also occasionally use.
This version of the map is appropriate for $\infty>v\geq1/2$, for which $e^{i\nu}$
is a pure phase. The case $v<1/2$ must be handled
separately. Some special values of $\alpha$ delineate the map: 
$\alpha=0$
corresponds to $e^{ik}\xi=1$ which implies $k=0$, and $\alpha=\pm\infty$
map to $k=\pm(\pi-2\nu)$. (We are choosing $k$ to be in the 
range $-\pi<k<\pi$.) Thus the whole range $-\infty<\alpha<\infty$
corresponds to $-(\pi-2\nu)<k<\pi-2\nu$. Note that $v\to\infty$
shrinks the range of $k$ to 0, whereas $v\to1/2$ represents the
maximum range. It is straightforward to work out
the following quantities in terms of the new variables: 
\begin{eqnarray}
\tan k&=&{\sin2\nu\sinh\alpha\over\cos\nu-\cos2\nu\cosh\alpha}\nonumber\\
{dk\over d\alpha}&=&{\sin3\nu\over2[\cosh\alpha-\cos3\nu]}
+{\sin\nu\over2[\cosh\alpha-\cos\nu]}\nonumber\\
t(k)&=&\left(\cos k+{1\over v}\sqrt{1-v^2\sin^2 k}\right)^2
={\cosh\alpha-\cos3\nu\over\cosh\alpha-\cos\nu}\nonumber\\
A&=&-{1-e^{\beta-\alpha-4i\nu}\over e^{\beta-\alpha}-e^{-4i\nu}}
\equiv-e^{i\theta(\alpha,\beta)}\nonumber\\
\theta(\alpha,\beta)&=&2\tan^{-1}\cot2\nu\tanh((\beta-\alpha)/2)
\end{eqnarray}
Using the last two equations, we can express the boundary conditions
in the alternative forms
\begin{eqnarray}
e^{iMk_l}&=&\prod_{j\neq l}A(j,l)=(-)^{q-1}e^{i\sum_{j\neq l}
\theta(\alpha_j,\alpha_l)}\nonumber\\
k_l&=&{2\pi I_l\over M}+{1\over M}\sum_{j\neq l}\theta(\alpha_j,\alpha_l),
\end{eqnarray}
where the $I_l$ are integers when $q$ is odd, and they are half-odd integers
when $q$ is even. Different choices for these integers lead
to different solutions for the set of $k$'s. Yang and Yang \cite{yangyang}
encounter similar equations in their analysis of the $x, y$
Heisenberg spin chain, and their techniques for solving them
in the limit $M\to\infty$ can be directly applied. For easy comparison,
we attempt as far as possible to adopt their notation. 
\subsection{Consecutive $I_l$: $Q,P\neq0$}
We begin by first choosing the set of numbers $I_l$ to be consecutive
with no gaps: $I_{l+1}=1+I_l$. We define a kernel $K$ and density function
$R(\alpha)$ by
\begin{eqnarray}
K(\alpha,\beta)&\equiv&{1\over2\pi}{\partial\theta\over\partial\beta}
={1\over2\pi}{\sin4\nu\over\cosh(\alpha-\beta)-\cos4\nu}\nonumber\\
R(\alpha)&=&{2\pi\over M}{dj\over d\alpha},
\end{eqnarray}
and then convert the equation for the $k$'s as $M\to\infty$ into
an integral equation
\begin{eqnarray}
{dk\over d\alpha}&=&R(\alpha)+\int_{\alpha_-}^{\alpha_+}{d\beta}
K(\alpha-\beta)R(\beta).
\end{eqnarray}
This equation has the same kernel $K$ as the one analyzed in \cite{yangyang},
 but a different inhomogeneous term $dk/d\alpha$.
The values chosen for $\alpha_\pm$ determine the characteristics of the
eigenstate. For example, the eigenstate with maximum eigenvalue
$T$ for the transfer matrix corresponds to $\alpha_\pm=\pm\infty$.
The values of $k$ at the limits of this range are $k=\pm(\pi-2\nu)$
and clearly $t(k)=1$ for these values. As long as $0<\nu<\pi/2$,
$t(k)>1$ for all finite $\alpha$, so taking the whole range of
$\alpha$ corresponds to including in the expression for $T$
all values for $t$ greater than unity. For the continuum limit
we are only interested in very large $\alpha_\pm$ since then
the eigenvalues will be close (within $1/M$) of the maximum
eigenvalue.

As shown in \cite{yangyang}, 
the kernel $J=-(I+K)^{-1}K$, can be used to rewrite the equation
for $R$, which determines it over the whole range of $\alpha$,
in terms of its values outside the range $(\alpha_-,\alpha_+)$.
This is useful since we are interested only in the
excited states close to the ground state corresponding to
$\alpha_\pm=\pm\infty$.
\begin{eqnarray}
R(\alpha)=R_0(\alpha)
-\left[\int_{-\infty}^{\alpha_-}+\int^{\infty}_{\alpha_+}\right]
J(\alpha-\beta)R(\beta)
\label{reqoutside}
\end{eqnarray}
where $R_0$ is the solution of the equation for $\alpha_\pm=\pm\infty$. 
It can be easily found by Fourier
transformation of the equation. From
\begin{eqnarray}
{dk\over d\alpha}&=&\int d\lambda e^{-i\lambda\alpha}{\sinh(\pi-2\nu)\lambda
\cosh\nu\lambda\over\sinh\pi\lambda}\nonumber\\
K(\alpha)&=&\int_{-\infty}^{\infty}{d\lambda\over2\pi}\ e^{-i\lambda\alpha}\
{\sinh(\pi-2\mu)\lambda\over\sinh\pi\lambda},
\end{eqnarray}
we determine
\begin{eqnarray}
R_0(\alpha)&=&\int_{-\infty}^{\infty}{d\lambda}\ e^{-i\lambda\alpha}\
{\cosh\nu\lambda\over2\cosh2\nu\lambda}\nonumber\\
&=&{\pi\over\mu\sqrt2}{\cosh(\pi\alpha/2\mu)\over\cosh(\pi\alpha/\mu)}.
\end{eqnarray} 
Recall that $\mu=2\nu$.

We can also easily express $J$ as a Fourier integral:
\begin{eqnarray}
J(\alpha)&=&-\int_{-\infty}^{\infty}{d\lambda\over2\pi}\ e^{-i\lambda\alpha}\
{\sinh(\pi-2\mu)\lambda\over2\sinh(\pi-\mu)\lambda\cosh\mu\lambda}.
\end{eqnarray}
The conserved quantities $Q=M-2q, P=\sum_jk_j$, 
the total charge and total momentum
respectively can be expressed, in the limit
$M\to\infty$, as integrals either inside or outside
the range $(\alpha_-,\alpha_+)$. These
expressions then implicitly determine $\alpha_\pm$ in terms of $Q, P$.
\begin{eqnarray}
{1\over2}-{Q\over2M}={q\over M}&=&
\int_{\alpha_-}^{\alpha_+}{d\beta\over2\pi}R(\beta)
\nonumber\\
&=&\int_{-\infty}^{\infty}{d\beta\over2\pi}R(\beta)
-\left[\int_{-\infty}^{\alpha_-}+\int^{\infty}_{\alpha_+}\right]
{d\beta\over2\pi}R(\beta)\nonumber\\
&=&{1\over2}-\left[\int_{-\infty}^{\alpha_-}+\int^{\infty}_{\alpha_+}\right]
{d\beta\over2\pi}R(\beta)\left(1+\int_{-\infty}^{\infty}d\alpha 
J(\alpha-\beta)\right)
\end{eqnarray}
Now, 
\begin{eqnarray}
1+\int_{-\infty}^{\infty}d\alpha J(\alpha-\beta)
&=&1-{\pi-2\mu\over2(\pi-\mu)}={\pi\over2(\pi-\mu)},
\end{eqnarray}
so we have
\begin{eqnarray}
{Q\over M}={\pi\over \pi-\mu}\left[\int_{-\infty}^{\alpha_-}
+\int^{\infty}_{\alpha_+}\right]{d\beta\over2\pi}R(\beta).
\end{eqnarray}
In a similar manner we can express the total momentum as
\begin{eqnarray}
{P\over M}={1\over M}\sum_{j=1}^q k_j&=&\int_{\alpha_-}^{\alpha_+}{d\beta\over2\pi}R(\beta)k(\beta)\nonumber\\
&=&\int_{-\infty}^{\infty}{d\beta\over2\pi}R(\beta)
k(\beta)
-\left[\int_{-\infty}^{\alpha_-}+\int^{\infty}_{\alpha_+}\right]
{d\beta\over2\pi}R(\beta)k(\beta)\nonumber\\
&=&{P_0\over M}-\left[\int_{-\infty}^{\alpha_-}+\int^{\infty}_{\alpha_+}\right]
{d\beta\over2\pi}R(\beta)\left(k(\beta)
+\int_{-\infty}^\infty d\alpha J(\alpha-\beta)k(\alpha)
\right)
\end{eqnarray}
We can infer the Fourier transform of $k(\alpha)$ from that of $dk/d\alpha$.
\begin{eqnarray}
{dk\over d\alpha}&=&\int d\lambda e^{-i\lambda\alpha}{\sinh(\pi-2\nu)\lambda
\cosh\nu\lambda\over\sinh\pi\lambda}\nonumber\\
k(\beta)&=&-{1\over2i}\int d\lambda e^{-i\lambda\beta}
{\sinh(\pi-2\nu)\lambda
\cosh\nu\lambda\over\sinh\pi\lambda}\left[
{1\over\lambda+i\epsilon}+{1\over\lambda-i\epsilon}\right]\nonumber\\
k(\beta)+\int d\alpha J(\alpha-\beta)k(\alpha)
&=&-{1\over2i}\int d\lambda e^{-i\lambda\beta}
{
\cosh\nu\lambda\over2\cosh2\nu\lambda}\left[
{1\over\lambda+i\epsilon}+{1\over\lambda-i\epsilon}\right]\nonumber\\
&\to& \pm{\pi\over2},\qquad {\rm for} \quad\beta\to\pm\infty.
\end{eqnarray}
Note that the $i\epsilon$ prescription is chosen so that 
$k(\pm\infty)=\pm(\pi-2\nu)$, as required by the mapping.
Finally, since $P_0=0$, we have for large $\alpha_+,\alpha_-$,
\begin{eqnarray}
{P\over M}&\approx&-{\pi\over2}\left[\int^{\infty}_{\alpha_+}
-\int_{-\infty}^{\alpha_-}
\right]{d\beta\over2\pi}R(\beta).
\end{eqnarray}

Finally, we manipulate the expression for the energy, proportional to
$-\ln T$, expressing it as an integral outside the interval 
$(\alpha_-,\alpha_+)$:
\begin{eqnarray}
{\ln T\over M}&=&\int_{\alpha_-}^{\alpha_+}{d\beta\over2\pi}R(\beta)
\ln\left[{\cosh\beta-\cos3\nu\over\cosh\beta-\cos\nu}\right]\nonumber\\
&=&\int_{-\infty}^{\infty}{d\beta\over2\pi}R(\beta)
\ln\left[{\cosh\beta-\cos3\nu\over\cosh\beta-\cos\nu}\right]
-\left[\int_{-\infty}^{\alpha_-}+\int^{\infty}_{\alpha_+}\right]
{d\beta\over2\pi}R(\beta)\ln\left[{\cosh\beta
-\cos3\nu\over\cosh\beta-\cos\nu}\right]\nonumber\\
&=&{\ln T_0\over M}-\left[\int_{-\infty}^{\alpha_-}+\int^{\infty}_{\alpha_+}\right]
{d\beta\over2\pi}R(\beta)\left(\ln t(\beta)
+\int_{-\infty}^\infty d\alpha J(\alpha-\beta)\ln t(\alpha)
\right),
\end{eqnarray}
where we have written $t(\alpha)$ as shorthand for $t(k(\alpha))$.
We use the Fourier transform of $\ln t(\alpha)$, given by
\begin{eqnarray}
\nonumber\\
\ln t(\alpha)&=&\ln\left[{\cosh\alpha-\cos3\nu\over\cosh\alpha-\cos\nu}\right]
=2\int d\lambda e^{-i\lambda\alpha}{\sinh(\pi-2\nu)\lambda
\sinh\nu\lambda\over\lambda\sinh\pi\lambda},
\end{eqnarray}
to arrive finally at a convenient expression for $E-E_0$
(Recall that $T=e^{-2aE}$, since the transfer matrix evolves
two discrete time steps.):
\begin{eqnarray}
{E-E_0\over M}&=&-{\ln T-\ln T_0\over2a M}\nonumber\\
&=&{1\over2a}\left[\int_{-\infty}^{\alpha_-}+\int^{\infty}_{\alpha_+}\right]
{d\beta\over2\pi}R(\beta)\int_{-\infty}^\infty d\lambda e^{-i\lambda\beta}
{\sinh\nu\lambda\over\lambda\cosh2\nu\lambda}\nonumber\\
&\approx&{1\over2a}\left[\int_{-\infty}^{\alpha_-}
+\int^{\infty}_{\alpha_+}\right]
{d\beta\over2\pi}R(\beta){\sqrt2\over\cosh(\pi\beta/2\mu)},
\end{eqnarray}
where the last line is approximate, assuming large $\alpha_+,\alpha_-$.
It is arrived at by first evaluating
\begin{eqnarray}
{\partial\over\partial\beta}\int_{-\infty}^\infty d\lambda e^{-i\lambda\beta}
{\sinh\nu\lambda\over\lambda\cosh2\nu\lambda}&=&
-i\int_{-\infty}^\infty d\lambda e^{-i\lambda\beta}
{\sinh\nu\lambda\over\cosh2\nu\lambda}\nonumber\\
&=&-{\pi\sqrt2\over\mu}{\sinh\pi\beta/2\mu\over\cosh\pi\beta/\mu}\nonumber\\
&\to&\cases{-\pi\sqrt2e^{-\pi\beta/2\mu}/\mu&for $\beta\to+\infty$\cr
+\pi\sqrt2e^{\pi\beta/2\mu}/\mu&for $\beta\to-\infty$\cr},
\end{eqnarray}
and then integrating the asymptotic form to get
\begin{eqnarray}
\int_{-\infty}^\infty d\lambda e^{-i\lambda\beta}
{\sinh\nu\lambda\over\lambda\cosh2\nu\lambda}
&\to&\cases{2\sqrt2e^{-\pi\beta/2\mu}&for $\beta\to+\infty$\cr
2\sqrt2e^{\pi\beta/2\mu}&for $\beta\to-\infty$\cr}\nonumber\\
&\approx& {\sqrt2\over\cosh(\pi\beta/2\mu)}.
\end{eqnarray}

To find the energy levels close to the ground state, we must 
analyze the equations for $R$ for large $\alpha_+, \alpha_-$.
For $\alpha>\alpha_+$ Eq.~\ref{reqoutside} can be approximated
by dropping the integral over negative $\alpha$ and using the
asymptotic form for $R_0$:
\begin{eqnarray}
R(\alpha)+\int^{\infty}_{\alpha_+}J(\alpha-\beta)R(\beta)&\approx&
{\pi\over\mu\sqrt2}e^{-\pi\alpha/2\mu}.
\label{req+}
\end{eqnarray}
It is convenient to put 
$$R(\alpha+\alpha^+)
={\pi\over\mu\sqrt2}e^{-\pi\alpha_+/2\mu}S(\alpha)$$
so that Eq.~\ref{req+} reduces to the Wiener-Hopf equation \cite{yangyang}
\begin{eqnarray}
S(\alpha)+\int^{\infty}_0 J(\alpha-\beta)S(\beta)&=&
e^{-\pi\alpha/2\mu}.
\label{wienerhopf}
\end{eqnarray}
Similarly, analyzing the equation for $\alpha<\alpha_-$, leads to
the identification
$$R(\alpha+\alpha^-)
\approx{\pi\over\mu\sqrt2}e^{\pi\alpha_-/2\mu}S(-\alpha).$$
Inserting these approximations into the formulas for $Q$, $P$,
and $E$, leads to 
\begin{eqnarray}
{Q\over M}&\approx&{\pi\over\pi-\mu}{1\over2\mu\sqrt2}
\left[e^{-\pi\alpha_+/2\mu}+e^{\pi\alpha_-/2\mu}\right]\int_0^\infty
d\beta S(\beta)\nonumber\\
{P\over M}&\approx&-{\pi\over2}{1\over2\mu\sqrt2}
\left[e^{-\pi\alpha_+/2\mu}-e^{\pi\alpha_-/2\mu}\right]\int_0^\infty
d\beta S(\beta)\nonumber\\
{E-E_0\over M}&\approx&{1\over2a\mu}
\left[e^{-\pi\alpha_+/\mu}+e^{\pi\alpha_-/\mu}\right]\int_0^\infty
d\beta S(\beta)e^{-\pi\beta/2\mu}.
\end{eqnarray}
Next one can solve the first two equations for $\alpha_+$ and
$\alpha_-$ and substitute in the last equation to get
\begin{eqnarray}
{E-E_0\over M}&\approx&{4\mu\over2a}{I(\pi/2\mu)\over I(0)^2}
\left[{(\pi-\mu)^2\over\pi^2}{Q^2\over M^2}+{4\over\pi^2}
{P^2\over M^2}\right],
\end{eqnarray}
where we have defined $I(x)=\int_0^\infty d\beta S(\beta)e^{-x\beta}$.
From the solution of Eq.~\ref{wienerhopf} one can infer (see
\cite{yangyang}) that 
${I(\pi/2\mu)/I(0)^2}=\pi^2/8\mu(\pi-\mu)$, so finally
\begin{eqnarray}
{E-E_0}&\approx&{1\over aM}
\left[{\pi-\mu\over4}{Q^2}+{1\over\pi-\mu}{P^2}\right]
={T_0\over P^+} 
\left[{\pi-\mu\over4}{Q^2}+{1\over\pi-\mu}{P^2}\right],
\label{gaplessexc}
\end{eqnarray}
regaining a key result of \cite{gilesmt}. 
\subsection{Non-consecutive $I_l$}
The excited states included
in Eq.~\ref{gaplessexc} are those where the numbers $I_l$
are consecutive. For example, the state with $Q=P=0$
corresponds to the choice (with $q=M/2$ odd)
$$\left(-{q-1\over2},\ldots,{q-3\over2},{q-1\over2}\right).$$
There are also excitations in which
``holes'' are allowed in this set of numbers. As an example,
consider replacing $(q-1-2j)/2$ in the above list by $(q+1)/2$,
creating a gap, but retaining the same number
of overturned arrows, so that $Q=0$. However the momentum
is increased by the amount $P=2\pi(j+1)/M$. For large $M$,
the effect of this hole on the $k$'s is small, and it
makes sense to expand them around the values appropriate
to the $Q=P=0$ state, the new set of $k$'s differing from the
latter by $\delta k_j$. Referring to the original equation
for the $k$'s, we find an equation for $\delta k$:
\begin{eqnarray}
\delta k_l&=&{2\pi\over M}\theta(l-l_j)+{2\pi \over M}
\sum_{j\neq l}\left[-{\partial\alpha_j\over\partial k_j}
\delta k_j+{\partial\alpha_l\over\partial k_l}
\delta k_l\right]K(\alpha_l-\alpha_j)\nonumber\\
\delta k_l\left(1-{\partial\alpha_l\over\partial k_l}
{2\pi\over M}\sum_{j\neq l}K(\alpha_l-\alpha_j)\right)
&=&{2\pi\over M}\theta(l-l_j)-{2\pi\over M}
\sum_{j\neq l}\left[{\partial\alpha_j\over\partial k_j}
\delta k_l\right]K(\alpha_l-\alpha_j)\nonumber\\
\delta k(\alpha){\partial\alpha\over\partial k}R(\alpha)
&=&{2\pi\over M}\theta(\alpha-\alpha_j)-
\int d\beta{\partial\beta\over\partial k}
\delta k(\beta)R(\beta)K(\alpha-\beta),
\end{eqnarray}
where we have replaced the sums by integrals in the last line.
Defining $\chi(\alpha)=M\delta k(\alpha)R(\alpha)d\alpha/dk$,
we have the integral equation
\begin{eqnarray}
\chi(\alpha)+\int_{-\infty}^\infty d\beta K(\alpha-\beta)\chi(\beta)
&=&2\pi\theta(\alpha-\alpha_j).
\label{pheq}
\end{eqnarray}
Here $\alpha_j$ marks the location of the ``hole''. It can be 
related to the value for the momentum of the excited state:
\begin{eqnarray}
P&=&{2\pi(j+1)\over M}=\sum_l \delta k_l
=\int_{-\infty}^\infty {d\alpha\over2\pi}\chi(\alpha)
{dk\over d\alpha}
\end{eqnarray}
Similarly, we can write the energy difference between the excited
and ground state as
\begin{eqnarray}
E-E_0={1\over2a}(\ln T_0-\ln T)&=&-{1\over2a}\sum_l\delta k_l
{d\alpha_l\over dk_l}{1\over t_l}{dt_l\over d\alpha_l}\nonumber\\
&=&-{1\over2a}\int_{-\infty}^\infty {d\alpha\over2\pi}\chi(\alpha)
\left[{\sinh\alpha\over\cosh\alpha-\cos3\nu}
-{\sinh\alpha\over\cosh\alpha-\cos\nu}\right]\nonumber\\
\end{eqnarray}
Eq.~\ref{pheq} can be immediately solved via Fourier transformation:
\begin{eqnarray}
\chi(\alpha)&=&i\int d\lambda e^{-i(\alpha-\alpha_j)\lambda}{\sinh\pi\lambda
\over2(\lambda+i\epsilon)\sinh(\pi-\mu)\lambda \cosh\mu\lambda},
\end{eqnarray}
and used to obtain the total momentum and energy 
\begin{eqnarray}
P&=&{i\over2}\int{d\lambda}\ {e^{-i\lambda\alpha_j}
}{\cosh\nu\lambda\over(-\lambda+i\epsilon)\cosh2\nu\lambda}\nonumber\\
E-E_0&=&-{1\over2a}\int{d\lambda}\ {e^{-i\lambda\alpha_j}
}{\sinh\nu\lambda\over(-\lambda+i\epsilon)\cosh2\nu\lambda}.
\end{eqnarray}
Of course, we are interested in these expressions in the limit
$\alpha_j\to\infty$, corresponding to the continuum limit. This
asymptotic limit is obtained by deforming the integration 
contours into the lower half plane and picking up the
nearest pole to the real axis, namely the one at $\lambda=-i\pi/4\nu$.
This leads to
\begin{eqnarray}
P\sim\sqrt2 e^{-\pi\alpha_j/4\nu} \qquad\qquad E-E_0\sim{\sqrt2\over a}
e^{-\pi\alpha_j/4\nu},
\end{eqnarray}
from which we conclude that
\begin{eqnarray}
E-E_0={P\over a}={2\pi(j+1)\over Ma}={2\pi(j+1)T_0\over P^+}
\end{eqnarray}
in the limit $M\to\infty$. Notice the important fact that the
energy of these excitations is independent of the vertex weight
$v$. Although we have discussed only one particular ``particle-hole''
excitation, it is clear that the energy of the state with many particle-hole
pairs will simply be additive in the momentum carried by each
pair. Furthermore, there are two independent
sets of such excitations about the two boundaries of the
Fermi sea. Each particle hole excitation contributes $2\pi nT_0/P^+$,
where $n>0$. If there are several particle-hole pairs from the right
side  $k>0$ of the Fermi sea, we define $N_R=\sum_i n_i$, and similarly $N_L$
is defined for those from the left side $k<0$ of the Fermi sea. 

These contributions to the energy are added to those arising
from non-zero $Q, P$.
Note that the $P^2$ term in the energy receives negligible contributions
from particle-hole excitations from the same side of the Fermi
sea, since these have $P=O(1/M)$. 
This term is non-zero in the continuum limit only if the particle and hole
are from opposite sides of the sea. For example, replacing
$-(q+1)/2$ with $(q+1)/2$ contributes $2\pi q/M\approx\pi$ to
$P$. But such large momentum pair excitations have already been accounted
for among the excitations with consecutive $I_l$ considered earlier.
Thus the energy levels of the continuum limit are determined
by $Q$, $P$, $N_R$, and $N_L$:
\begin{eqnarray}
E-E_0&=&{T_0\over2P^+}\left[{\pi-\mu\over2}{Q^2}+{2\over\pi-\mu}{P^2}
+{4\pi(N_R+N_L)}\right].
\end{eqnarray}
Recall that $Q=2r$ and $P=\pi s$ where $r, s$ range independently over all
integers.

\section{Discussion and Concluding Remarks}
We conclude by recalling the comparison of the energy spectrum
just obtained with that of a compactified scalar field on
the continuum closed string world-sheet \cite{gilesmt},
described by the action
\begin{eqnarray}
S&=&{1\over2}\int d\tau\int_0^{P^+}d\sigma({\dot\phi}^2-T_0^2{\phi^\prime}^2)
\end{eqnarray}
with the equivalence relation
\begin{eqnarray}
\phi&\equiv& \phi+2\pi R.
\end{eqnarray}
We would like to identify $\phi/R$ with the angle of the $O(2)$
rotations of the underlying symmetry of the theory. It should therefore
be conjugate to the charge $Q$. Thus we take $Q$ proportional
to the zero mode of the conjugate momentum to $\phi$,
\begin{eqnarray}
\int_0^{P^+}d\sigma{\dot\phi}={k\over R}={Q\over2R}.
\end{eqnarray}
Then $P$ must be taken proportional to the winding number $l$
defined by the boundary condition $\phi(P^+)=\phi(0)+2\pi lR$. 
The energy of the compactified scalar field is therefore
\begin{eqnarray}
E={1\over2P^+}\left[{k^2\over R^2}+{4\pi^2l^2T_0^2R^2}
+{4\pi T_0(N_R+N_L)}\right],
\end{eqnarray}
from which we infer that $R^2=1/(2T_0(\pi-\mu))$.
Remembering that $\cos\mu=\cos2\nu={\rm Re}~e^{2i\nu}=-1+1/2v^2$,
we see that the radius of the circle on which the field
lives is determined by the ratio of quartic couplings. In
particular the limit $R\to\infty$ implies $\mu\to\pi$ or $v\to\infty$.
The self dual radius $R_*^2=1/2\pi T_0$ corresponds to $\mu=0$
or $v=1/2$. Thus the range of couplings considered here
$1/2\leq v<\infty$ (for which the 6-vertex model is
critical) produces circle radii $R_*\leq R <\infty$.
Interestingly, small radii, $R<R_*$ are not accessible
in the vertex model. For $v<1/2$ the model is not critical
and the continuum limit accordingly sends all excitations
to infinite energy, i.e. there is no interesting continuum
limit.
 
The important message of \cite{gilesmt} is that if string can
be understood as a composite of field quanta along the
lines of the fishnet model, then internal degrees of freedom
carried by the fields are naturally promoted to world-sheet
fields. Among the possible interpretations of these world-sheet
fields is that they represent extra compact dimensions.
The simple $O(2)$ model reviewed in this article leads to the
emergence of an extra circular dimension $S_1$. Within the fishnet model, 
it will certainly be interesting to study the $SO(n)$ case. If
the $n=2$ case is a fair guide, we can hope that what will
emerge is a world-sheet field that lives on the sphere $S_{n-1}$.
This would be the $SO(n)$ nonlinear sigma model.
The vertex model in this case is much more complicated
than the 6-vertex model: each line of the diagram can be in $n$
states, and there are many more distinct vertices. For instance,
in the $n=3$ case there are 19 vertices that conserve the 3-component
of spin. These models are currently under study. 

The fishnet model is at best relevant only to the strong
coupling behavior of the underlying quantum field theory. Even
then we do not yet fully understand how fishnets fit in to the most
well-established string/field duality, 
AdS$_5\times$S$_5$/${\cal N}=4$~SUSY Yang-Mills. Optimistically,
the mechanism illustrated in this paper will shed light on
the field theoretic origin of the S$_5$ part of the ten dimensional
space-time manifold: the Yang-Mills super-multiplet contains
6 scalar fields that transform under the vector representation
of $SO(6)$. The conjecture of the previous paragraph suggests that
fishnets propagating only these scalars would describe the $SO(6)$
nonlinear sigma model, leading to the emergence of $S_5$. 

But in its simplest presentation, the dense fishnet model
produces flat Minkowski space-time for the remaining non-compact
dimensions, not anti-de Sitter space-time.
We believe that this flaw is present because   
the dense fishnets completely freeze out both fluctuations in the number of
field quanta {\it and} fluctuations in the distribution of $P^+$
along string.
Properly including the gauge fields in the fishnet approach  
could improve this situation by allowing some of these
fluctuations even at strong coupling \cite{beringrt}. Indeed,
since the $SO(n)$ nonlinear sigma model is not conformally
invariant for $n>2$, consistency of the string interpretation requires a 
nontrivial dynamical interplay between the S$_5$ degrees of freedom and
the space-time degrees of freedom in the world-sheet action. 
As presently understood, the AdS/CFT connection \cite{maldacena}
implements this via the AdS metric and the 5-form Ramond-Ramond
flux on $S_5$.
More concretely, the light-cone treatment of string on AdS$_5\times$S$_5$
\cite{metsaevt,metsaevtt} indicates how these interactions
appear in the world-sheet action, in which the 5th dimension  
plays a role similar to the Liouville field in restoring
conformal invariance. It remains to be seen whether the graph
summation approach can adequately account for these features.

\smallskip
\underline{Acknowledgments:} I wish to thank Arkady Tseytlin for
many stimulating discussions and Achim Kempf for critically
reading the manuscript. This work is supported in part by U.S. DOE grant 
DE-FG02-97ER-41029.


\end{document}